\documentstyle[preprint,aps,prb,amstex]{revtex}
\begin{document}

\draft
\flushbottom

\title{Defects as a reason of continuity of normal-incommensurate phase transitions}
\author{A. Cano and A.P. Levanyuk}
%EndAName
%\textit{Departamento de F\'\i sica de la Materia Condensada.}\\
%\textit{Universidad Aut\'onoma de Madrid, 28049 Madrid, Spain}}

\address{Departamento de F\'\i sica de la Materia Condensada.
Universidad Aut\'onoma de Madrid, 28049 Madrid, Spain}
\date{\today}
\maketitle

\begin{abstract}
Almost all normal-incommensurate phase transitions observed experimentally are
continuous. We show that there is not any theoretical reason for this
general behaviour in perfect crystals. A normal-incommensurate phase transition that is not too
far from the mean-field tricritical point should be discontinuous and it is
highly improbable that so far reported normal-incommensurate phase
transitions lie very far from this point. To understand
this behaviour we study influence of defects on a hypothetical first-order
normal-incommensurate phase transition in a pure material. We have found
that this influence is strikingly different from that on other kinds of
first-order phase transitions. The change of the discontinuity of the order
parameter at the transition is negative and formally diverges within our
approximate theory. At the same time the diminishing of the phase transition
temperature remains finite. We interpret these results as an indication that at least some of the observed seemingly second-order normal-incommensurate transitions would be first-order transitions in defectless crystals.
\end{abstract}

\pacs{PACS numbers: 61.44.Fw, 64.60.-i}

\section{Introduction}

Almost all observed structural normal-incommensurate (N-IC) phase transitions are
continuous (see for instance Ref.\onlinecite{blinclibro} as example of continuous cases and Ref.\onlinecite{lukianchuk} for rare examples of discontinuous ones). This is in contrast with other types of
structural phase transitions that are generally discontinuous. Such a
discontinuity is not surprising because, as was pointed out by many authors
\cite{larkin69}, the second-order (continuous) phase transitions convert
into first-order (discontinuous) ones because of many factors like shear
deformations, crystal anisotropy, etc..., except for phase transitions involving uniaxial ferroelectrics
and ferroelastics. These factors are operative by virtue of order parameter
fluctuations. We show that also a broad class of normal-incommensurate
(N-IC) phase transitions should be discontinuous in pure materials due to
the solid-state elasticity. We refer to transitions that are
not too far from the mean-field tricritical point. Then a question arises:
why all observed experimentally N-IC phase transition are continuous?

In perfect crystals the N-IC
first-order phase transitions should exist but it is reasonable to assume that in real crystals the defects can smear them \cite{Lev99}. Indeed, it is well known \cite{larkin} that some defects influence dramatically on the incommensurate phases, destroying the long-range order at any concentration. Therefore strictly speaking, there are no N-IC phase transitions in real crystals or in other words, these 
transitions are always smeared. However, this smearing is not observed 
in the experiments and N-IC phase transitions appears usually as "ordinary" transitions, i.e. without apparent smearing in the measured anomalies. As we have mentioned, what is surprising is that almost all N-IC phase transitions are second-order ones. We try to understand this general behaviour showing that first-order N-IC transitions
can assume a continuous appearance if the crystal contains defects.

The paper is organized as follows. In Sec.II we demonstrate that close to the
mean-field tricritical point a second-order N-IC phase transition converts
into a first-order one because of the combined effect of the solid-state elasticity and
the order parameter fluctuations. It is important that the conversion takes
place even in the framework of the first order perturbation theory of effects of critical fluctuations.
The vicinity of the tricritical point where this conversion takes place is
shown to be broad. In Sec.III we present the method to
consider influence of defects on first-order phase
transitions\cite{Lev97} \ \cite{Lev98} within first
order perturbation theory where only linear terms in defect concentration
taken into account (see e.g. Ref.\onlinecite{libro2,Lev88}). To simplify
the treatment, the phase
transition in the pure material is supposed to be describable within the
classical Landau theory. In Sec.IV we apply the
method to a first-order N-IC phase transition. Because of the specific
features of a incommensurate phase, this theory is not rigorously
applicable here but we use the fact that N-IC transitions can be considered
as a limiting case of some normal-commensurate (N-C) transitions. So we consider
influence of defects on a normal-commensurate (N-C) first-order phase
transition where the perturbation theory is valid, and then trace the
changes produced by defects when the phase transition approaches a N-IC one.
As we will see, we cannot reach the N-IC phase transition because that means to go beyond the applicability of the perturbation theory, but
we can reveal some qualitative trends which provide us a base for the
speculations. In Sec.V we show that the same trends take place when the
first-order phase transition in the pure material is due to the combined
effect of order parameter
fluctuations and the solid-state elasticity as in
Sec.II. Finally, the Sec.VI is devoted to the conclusions.

\section{Fluctuation-induced first-order N-IC phase transitions}

Among several factors that are relevant in the conversion of a second-order
phase transition into a first-order one the most important one seems to be the
solid-state elasticity. Indeed, as it is shown in Ref.\onlinecite{bruno80},
close to the mean-field tricritical point the combined effect of the
order parameter fluctuations and the solid-state elasticity is so strong that
the second-order phase transition converts into a first-order one quite
far away from the region where the critical fluctuations become too much
important. In other words, this converted phase transition takes place in the region of applicability of
the Landau theory where the effects of the critical fluctuations can be
considered within the first order perturbation theory. This question was
reconsidered for structural displacive transitions in Ref.\onlinecite{Lev93}, and
it was shown that the region where the above-mentioned mechanism of conversion of a
second-order into first-order phase transition results operative is very broad.
Only the case of one component order parameter was considered. We shall show
now how the same conclusions are valid for a N-IC phase transition.

We restrict ourselves to the case of one-${\mathbf k}$ incommensurate phase,
i.e. the case of two component order parameter (designated as ${\boldsymbol
\eta ({\mathbf r})}=(\eta_1({\mathbf r}),\eta_2({\mathbf r}))$) \cite{libro1}. Then, the Landau continuous-media
potential takes the form: 
\begin{equation}
\Phi ({\boldsymbol \eta },u_{ik})=\int \left[ \varphi_{{\boldsymbol \eta
}}({\mathbf %
\eta })+\varphi_u({\boldsymbol \eta },u_{ik})\right] dv
\end{equation}
with 
\begin{eqnarray}
\varphi _{{\boldsymbol \eta }}({\boldsymbol \eta })=\frac 12A(\eta _1^2({\mathbf r}%
)+\eta _2^2({\mathbf r}))+\frac 14B(\eta _1^2({\mathbf r})+\eta_2^2({\mathbf r}%
))^2+ \nonumber\\ +\frac 16C(\eta _1^2({\mathbf r})+\eta _2^2({\mathbf r}))^3+\frac 12D((%
{\mathbf \nabla }\eta _1({\mathbf r}))^2+({\mathbf \nabla }\eta _2({\mathbf r}%
))^2)  \label{pot1}
\end{eqnarray}
and 
\begin{equation}
\varphi _u({\boldsymbol \eta },u)=r(\eta _1^2({\mathbf r})+\eta _2^2({\mathbf r}%
))u_{ll}({\mathbf r})+\frac 12Ku_{ll}^2({\mathbf r})+\mu (u_{ik}({\mathbf r}%
)-\frac 13u_{ll}({\mathbf r})\delta _{ik})^2
\end{equation}
where $K$ is the bulk modulus and $\mu $ is the shear modulus (to compare
with Ref.\onlinecite{Lev93}). We shall assume that the coefficient $A$
in Eq. \ref{pot1} is the only temperature dependent coefficient:
$A=\alpha(T-T_c)$
where $\alpha $ is a positive constant and $T_c$ would be the second-order transition temperature if coupling between elasticity and fluctuations is neglected. The other coefficients may depend on pressure but we consider
them positive constants.

As usually, it is assumed that the Landau potential is obtained as result
of a partial integration of the partition function over all degrees of
freedom, except those that correspond to the long-wave Fourier components of $%
{\boldsymbol \eta }({\mathbf r})$ and $u_{ik}({\mathbf r})$. To find thermodynamic
quantities, one has to integrate besides all the
degrees of freedom except those probed in experiment. For our purpose it is enough to leave unintegrated the zero
Fourier component of the order parameter. As in Ref.\onlinecite{Lev93} we minimize over the elastic
degrees of freedom. At this point one has to make out the spatially
homogeneous ($u_{ij}^{(0)}$) and inhomogeneous ($u_{ij}^{(i)}$) parts of
strain: 
\begin{equation}
u_{ij}({\mathbf r})=u_{ij}^{(0)}+u_{ij}^{(i)}=u_{ij}^{(0)}+\frac {1}{2}i{\sum_{{\mathbf k}\neq 0} }\left[
k_iu^j({\mathbf k})+k_ju^i({\mathbf k})\right] e^{i{\mathbf k\cdot r}}
\end{equation}
where ${\mathbf u}({\mathbf k})$ is ${\mathbf k}$-Fourier component of the
displacement vector. Minimizing over the homogeneous and the inhomogeneous
strain separately one obtains: 
\begin{eqnarray}
\int \varphi _u({\boldsymbol \eta },u)dv &=&-\frac{r^2}{2K}\left( {%
\sum_{{\mathbf k}} }\left[ \eta _{1{\mathbf k}}\eta _{1{\mathbf -k}}+\eta _{2{\mathbf
k}}\eta _{2-{\mathbf k}}\right] \right) ^2- 
\nonumber \\
&&-\frac{r^2}{2\gamma }{\sum_{{\mathbf k}\neq 0} }{\sum_{{\mathbf
k}_1,{\mathbf k}_2} }%
\left[ \eta _{1{\mathbf k}_1}\eta _{1-{\mathbf k}_1-{\mathbf k}}+\eta
_{2{\mathbf k}_1}\eta _{2-{\mathbf k}_1-{\mathbf k}}\right] \left[
\eta _{1-{\mathbf k}_2}\eta _{1{\mathbf k}_2+{\mathbf k}}+\eta _{2-{\mathbf
k}_2}\eta _{2{\mathbf k}_2+{\mathbf k}}\right]
\end{eqnarray}
where $\gamma =K+\frac 43\mu $ and we put the system volume equal to unity.

Taking these terms into account, the Landau potential can be
rewritten (limiting us to harmonic terms) as: 
\begin{equation}
\Phi ({\boldsymbol \eta }_0,...,{\boldsymbol \eta}_{\mathbf k},...)=\Phi _0({\boldsymbol \eta }_0)+\Phi _h({\boldsymbol \eta }_0,%
{\boldsymbol \eta }_{{\mathbf k}\neq 0})
\end{equation}
where $\Phi _0$ is the Landau potential that depends only on the zero
Fourier components of ${\boldsymbol \eta }$ (designated for simplicity as $%
{\boldsymbol \eta }_0=(\eta _1,\eta _2)$): 
\begin{equation}
\Phi _0({\boldsymbol \eta }_0)=\frac 12A(\eta _1^2+\eta _2^2)+\frac 14\widetilde{B%
}(\eta _1^2+\eta _2^2)^2+\frac 16C(\eta _1^2+\eta _2^2)^3
\end{equation}
and $\Phi _h$ are the harmonic terms of the Landau potential that depend on $%
{\boldsymbol \eta }_0$ and ${\boldsymbol \eta }_{{\mathbf k}\neq 0}$: 
\begin{equation}
\Phi _h({\boldsymbol \eta }_0,{\boldsymbol \eta }_{{\mathbf k}\neq
0})={\sum_{{\mathbf k}\neq 0}}
\frac 12\eta _{i{\mathbf k}}M_{ij}\eta _{j-{\mathbf k}}
\end{equation}
where \begin{equation}
M_{11}=A+3B_1\eta _1^2+\widetilde{B}\eta _2^2+C(5\eta _1^4+\eta _2^4+6\eta
_1^2\eta _2^2)+Dk^2
\end{equation}\begin{equation}
M_{22}=A+3B_1\eta _2^2+\widetilde{B}\eta _1^2+C(5\eta _2^4+\eta _1^4+6\eta
_1^2\eta _2^2)+Dk^2
\end{equation}\begin{equation}
M_{12}=(2\widetilde {B}+3\Delta )\eta _1\eta _2+4C(\eta _1\eta _2^3+\eta
_1^3\eta _2)
\end{equation}
In the preceding equations the constants $\widetilde{B},$ $B_1$ are
defined as: 
\begin{equation}
\widetilde{B}\equiv B-\frac{2r^2}K
\end{equation}
\begin{equation}
B_1\equiv B-\frac{2r^2}{3K}-\frac{4r^2}{3\gamma }=\widetilde{B}+%
\frac{16r^2\mu }{9K\gamma }\equiv \widetilde{B}+\Delta 
\end{equation}

As we have mentioned, we must integrate over the degrees of freedom corresponding
to the non-zero Fourier components of the order parameter according to the
Gibbs distribution. Thus, the incomplete thermodynamic potential is: 
\begin{equation}
\Phi =\Phi _0({\boldsymbol \eta }_0)-T\ln \left\{ \int \exp \left( -\frac{\Phi _h(%
{\boldsymbol \eta }_0,{\boldsymbol \eta }_{{\mathbf k}\neq 0})}T\right)
{\prod_{{\mathbf k}\neq 0} }
d\eta _{1{\mathbf k}}d\eta _{2{\mathbf k}}\right\}   \label{integration}
\end{equation}
As $\Phi _h$ is quadratic in $\eta _{1k}$ and $\eta _{2k}$%
, the integration can be easily performed, obtaining: 
\begin{equation}
\Phi ({\boldsymbol \eta }_0)=\Phi _0({\boldsymbol \eta
}_0)+{\frac {T}{2}}\left\{{\sum_{{\mathbf k}} }
\ln \left( M_{11}M_{22}-M_{22}^2\right) \right\}   \label{result}
\end{equation}
Using the fact that in the incommensurate phase, the system is invariant
respect to the phase of the order parameter vector ${\boldsymbol \eta }$, we can
choose it freely. Taking $\eta _1\neq 0,$ $\eta _2=0$ we considerably simplify the expression of the thermodynamic potential because in this
case $M_{22}=0$. Then, carrying out the integration of the second term of
Eq. \ref
{result}, and bearing in mind that $Dk_{\max }^2\gg A+3B_1\eta
_1^2+5C\eta _1^4$, we obtain: 
\begin{eqnarray}
\widetilde{\Phi }({\boldsymbol \eta }_0) &=&\frac 12\left[ A+\frac{k_{\max
}T}{2\pi ^2D}(3B_1+\widetilde{B})\right] \eta _1^2+\frac 14\left[ \widetilde{B}+%
\frac{k_{\max }T}{\pi ^2D}6C\right] \eta _1^4+\frac 16C\eta _1^6-  \nonumber
\\ && -\frac T{12\pi D^{3/2}}\left\{ (A+3B_1\eta _1^2+5C\eta
_1^4)^{3/2}+(A+\widetilde{B}\eta _1^2+C\eta _1^4)^{3/2}\right\}\label{uctuaciones} 
\end{eqnarray}
There is a trivial renormalization of the reference energy taken into
account and denoted by the symbol $\widetilde{\Phi }$ in the left side of
the Eq. \ref{uctuaciones}. From the first two terms, one can see that the
fluctuations provoke a
renormalization also in the coefficients $A$ and $\widetilde{B}$. This
renormalization also takes place in the rest of the terms because of higher order corrections. Since
this renormalization is not of interest within our phenomenological theory,
henceforth that all the coefficients correspond to their
renormalizated values. Then, the final form of the thermodynamic potential
including the effects of order parameter fluctuations and solid-state
elasticity is: 
\begin{eqnarray}
\widetilde{\Phi }({\boldsymbol \eta }_0) &=&\frac 12A\eta _1^2+\frac 14\widetilde{%
B}\eta _1^4+\frac 16C\eta _1^6-  \nonumber \\
&&\ \ \ \ \ -m\left\{ (A+3B_1\eta _1^2+5C\eta _1^4)^{3/2}+(A+\widetilde{B}%
\eta _1^2+C\eta _1^4)^{3/2}\right\}   \label{fluctuaciones}
\end{eqnarray}
where $m\equiv T/12\pi D^{3/2}$.

The same discussion as in Ref.\onlinecite{Lev93} about the final character of the
phase transition can be carried out now for a N-IC one. As in that work we can see
that close to the mean-field tricritical point, the order parameter at the
transition temperature is such that: 
\begin{equation}
B_1\eta _1^2\gg A,\text{ }C\eta _1^4
\end{equation}
As in Ref.\onlinecite{Lev93} we find the value of the coefficient A at the
transition temperature ($A^{\prime}$) and the order parameter discontinuity at the
transition ($\eta_1^{\prime2}$): 
\begin{equation}
A^{\prime}=\frac{(3m)^{4/3}}{2(2C)^{1/3}}(3B_1 )^2
\end{equation}
\begin{equation}
\eta _1^{\prime2}=\left( \frac{3m}{2C}\right) ^{2/3}(3B_1 )
\end{equation}
We must conclude that the combined effect of order parameter fluctuations
and the solid-state elasticity over a N-IC phase transition forces it to be discontinuous. This assertion is correct in a region close to
the mean-field tricritical point, i.e. as can be see in Ref.\onlinecite{Lev93} in
a region such that $\widetilde{B}<\mu B/K$. It is highly unlikely that all
the reported N-IC phase transitions lie out of this region because it is very
broad for structural systems. Such a conclusion is only applicable when
dealing with perfect crystals because in our treatment we have forgotten about the
influence of defects.

\section{Influence of defects: the method}

In this section we expose briefly the simplest version of
treatment of influence of defects on first-order phase transitions \cite
{Lev98}. This allows us to present formulas that are easy to generalize in
the cases of our interest. What is more important: in these formulas we keep a term that, although
omitted in previous papers, it is essential in the context of this work.

Within this section we consider one component order parameter ($\eta $) and ''random local
field'' (RLF) defects \cite{libro2,Lev88,nat88}. The Landau
continuous-media thermodynamic potential without defects has the form:
\begin{equation}
\Phi =\int \varphi (\eta )dv
\end{equation}
with
\begin{equation}
\varphi (\eta )=\frac 12A\eta ^2({\mathbf r})+\frac 14B\eta ^4({\mathbf r}%
)+\frac 16C\eta ^6({\mathbf r})+\frac D2({\mathbf \nabla }\eta ({\mathbf r}))^2
\label{Landau}
\end{equation}
where $A=\alpha (T-T_s)$, $T_s$ is the temperature at the spinode of the
symmetrical phase and all other coefficients are supposed to be
temperature-independent. As we are considering a first-order transition we assume
that $B<0$ and all the other coefficients positive.

The first-order transition temperature and the value of the order parameter
discontinuity at the transition is obtained with ${\mathbf \nabla }\eta (%
{\mathbf r})=0$ and using the equilibrium equations: $\partial \Phi /\partial
\eta =0$ (minimum of thermodynamic potential condition) and $\Phi =0$
(equality of thermodynamic potentials of the two phases condition). One has:
\begin{equation}
\widetilde{A}+B\widetilde{\eta }^2+C\widetilde{\eta }^4=0  \label{minimo}
\end{equation}
\begin{equation}
\widetilde{A}\widetilde{\eta }^2+\frac 12B\widetilde{\eta }^4+\frac 13C%
\widetilde{\eta }^6=0  \label{continuidad}
\end{equation}
where $\widetilde{A}$ and $\widetilde{\eta }^2$ are the values corresponding
to the first-order phase transition temperature. From these
equations: 
\begin{eqnarray}
\widetilde{A} &=&\frac{3B^2}{16C}  \label{basicoa} \\
\widetilde{\eta }^2 &=&-\frac{3B}{4C}  \label{basicoeta}
\end{eqnarray}
in a perfect crystal.

To take into account RLF defects one has to add to the thermodynamic
potential density $\varphi (\eta )$ in Eq. \ref{Landau} a term:
\begin{equation}
-h({\mathbf r})\eta ({\mathbf r})
\end{equation}
where $h({\mathbf r})$ is a ''random local field'' in the form:
\begin{equation}
h({\mathbf r})=\sum\limits_jh_j\delta ({\mathbf r}-{\mathbf r}_j)
\end{equation}
where $h_j=\pm H$, $H$ is a coefficient characterizing the ''strength'' of
the defect localized at ${\mathbf r}={\mathbf r}_j$, and the sign of $h_j$ is
random.

The order parameter decomposition into Fourier ($\eta _{{\mathbf k}}$)
components is the basic tool to study the influence of defects, so
expressing:
\begin{equation}
\eta ({\mathbf r)=}\sum\limits_{{\mathbf k}}\eta _{{\mathbf k}}e^{i{\mathbf
k\cdot r}}
\end{equation}
we consider the thermodynamic potential $\Phi $ as a function of all
Fourier components. The next step is to minimize this function over all
Fourier components except for the zero component (${{\mathbf k}=0,}$ designated
by $\eta _0$) that at this stage is consider as a fixed parameter. One
obtains to the linear approximation:
\begin{equation}
\eta _{{\mathbf k}\neq 0}^{(eq)}=\frac{h_{\mathbf k}}{A+3B\eta _0^2+5C\eta _0^4+Dk^2}%\equiv \chi(\eta _0,k)h_{\mathbf k}
\end{equation}
Then, we use these equilibrium values in the thermodynamic potential and average over the
defect positions obtaining \cite{Lev97} \ \cite{Lev98}:
\begin{equation}
\Phi (\eta _0)=\frac 12A\eta _0^2+\frac 14B\eta _0^4+\frac 16C\eta _0^6+\frac{%
NH^2}{8\pi D^{3/2}}(A+3B\eta _0^2+5C\eta _0^4)^{1/2}  \label{termpotonecom}
\end{equation}
where $N$ is the defect concentration. Instead of Eqs. \ref{minimo} and \ref
{continuidad} that give to us the values of the order parameter
discontinuity at the transition and the transition temperature, one has now:
\begin{equation}
\widetilde{A}+B\widetilde{\eta }_0^2+C\widetilde{\eta }_0^4+n\frac{3B+10C%
\widetilde{\eta }_0^2}{(\widetilde{A}+3B\widetilde{\eta }_0^2+5C\widetilde{%
\eta }_0^4)^{1/2}}=0  \label{minimodefect}
\end{equation}
\begin{equation}
\widetilde{A}\widetilde{\eta }_0^2+\frac 12B\widetilde{\eta }_0^4+\frac 13C%
\widetilde{\eta }_0^6+2n(\widetilde{A}+3B\widetilde{\eta }_0^2+5C\widetilde{%
\eta }_0^4)^{1/2}=2n\widetilde{A}^{1/2}  \label{continuidadefect}
\end{equation}
where $n=NH^2/8\pi D^{3/2}$, $\widetilde{A}$ and $\widetilde{\eta }_0^2$ are
the values at the first-order phase transition temperature.
Let us mention that in previous papers \cite{Lev99,Lev97,Lev98} the right hand side of Eq. \ref{continuidadefect} corresponding to
thermodynamic potential of the symmetrical phase was set equal to zero. This
term would not really modify the previous papers results but plays a basic
role for a N-IC phase transition.

The applicability conditions of these formulas at first-order phase
transition temperature can be obtained in the same way as in Ref. \onlinecite
{libro2,Lev88} and implies for the symmetric and non-symmetric
phases: 
\begin{equation}
n\frac{C^{3/2}}{B^2}<1  \label{crit1}
\end{equation}
where the numerical factor (of magnitude order of unity) has been omitted.

Designating the changes produced by defects in $\widetilde{A}$ and $%
\widetilde{\eta }_0^2$ (Eqs. \ref{basicoa} and \ref{basicoeta}) as $\delta 
\widetilde{A}$ and $\delta \widetilde{\eta }_0^2$ one can linearize Eqs. \ref
{minimodefect} and \ref{continuidadefect} with respect to this values to
obtain:
\begin{equation}
\delta \widetilde{A}=-2n\sqrt{\frac C3}  \label{deltaA1}
\end{equation}
\begin{equation}
\delta \widetilde{\eta }_0^2=-14n\sqrt{\frac C{3B^2}}  \label{deltaeta1}
\end{equation}
One sees that RLF defects provoke both lowering of the phase transition
temperature and diminishing of the discontinuity of the order parameter at
the transition \cite{nota}. According to Eq. \ref{crit1} we cannot set $B=0$ in the last
equation. Moreover this linear approximation to Eqs. \ref{minimodefect} and 
\ref{continuidadefect} lose their applicability at the same time that the
equations do. So nonlinear corrections to equations \ref{deltaA1} and \ref
{deltaeta1} would be inconsistent with the applicability criterion for
equations \ref{minimodefect} and \ref{continuidadefect}.

\section{Approaching a Normal-Incommensurate Transition: defects and Landau
Theory}

In as follows we consider N-IC phase transition as a
special case of phase transition with two component order parameter ($\eta
_1 $ and $\eta _2$) corresponding to isotropy in order parameter space. This
means that all invariants that we must consider in Landau thermodynamics
potential are power of the second order invariant \cite{libro1}. In other
words, taking the Landau potential density in the form : 
\begin{eqnarray}
\varphi ({\boldsymbol \eta }) &=&\frac 12A(\eta _1^2({\mathbf r})+\eta _2^2(%
{\mathbf r}))+\frac 14B_1(\eta _1^2({\mathbf r})+\eta _2^2({\mathbf r}%
))^2+\frac 12B_2\eta _1^2({\mathbf r})\eta _2^2({\mathbf r})+\nonumber \\
&&+\frac 16C(\eta
_1^2({\mathbf r})+\eta _2^2({\mathbf r}))^3+  \frac 12D(({\mathbf \nabla }\eta _1({\mathbf r}))^2+({\mathbf \nabla }\eta
_2({\mathbf r}))^2)
\end{eqnarray}
we approach a N-IC first-order transition if $B_2\rightarrow 0$ and all
other coefficients are considered, positive except $B_1$ and temperature
independent except $A=\alpha (T-T_s)$. Now to make allowances for defects we
add to the thermodynamic potential density a term: 
\begin{equation}
-h_1({\mathbf r})\eta _1({\mathbf r})-h_2({\mathbf r})\eta _2({\mathbf r})
\end{equation}
where $h_1$ and $h_2$ are the two components of a ''random local field'': 
\begin{equation}
h_{1,2}({\mathbf r})=\sum\limits_jh_{1,2}^j\delta ({\mathbf r-r}_j)
\end{equation}
We assume that $\langle (h_1^j)^2\rangle =\langle (h_2^j)^2\rangle =H^2$ and
that the ''direction'' of the ''vector'' ${\mathbf h=(}h_1,h_2{\mathbf )}$ is
random, i.e. $\langle h_1^jh_2^j\rangle =0$, where $\langle $ $\rangle $
designates the average.

We apply the method of the preceding section to find out the
thermodynamic potential as a function of the zero Fourier components
(${\mathbf k}=0,$
designated for simplicity by $\eta _1$ and $\eta _2$) . It is convenient to suppose that $B_2>0$. The same results can be obtained for the case $B_2<0$ but the expressions are considerably longer in the latter case. Thus in what follows we take the case in what expressions are straightforward, i.e. $B_2>0$. Then, non symmetrical phase equilibrium state
corresponds to $\eta _1\neq 0,$ $\eta _2=0$ or vice versa. Taking $\eta _2=0$
we obtain instead of equation \ref{termpotonecom} for a one component order
parameter, the thermodynamic potential in the form: 
\begin{eqnarray}
\Phi (\eta _1,\eta _2) &=&\frac 12A\eta _1^2+\frac 14B_1\eta _1^4+\frac
16C\eta _1^6+  \nonumber \\
&&+n\left\{ (A+3B_1\eta _1^2+5C\eta _1^4)^{1/2}+(A+(B_1+B_2)\eta _1^2+C\eta
_1^4)^{1/2}\right\}
\end{eqnarray}
and instead of equilibrium equations \ref{minimo} and \ref{continuidad} : 
\begin{equation}
\widetilde{A}+B_1\widetilde{\eta }_1^2+C\widetilde{\eta }_1^4+n\left\{ \frac{%
3B_1+10C\widetilde{\eta }_1^2}{(\widetilde{A}+3B_1\widetilde{\eta }_1^2+5C%
\widetilde{\eta }_1^4)^{1/2}}+\frac{(B_1+B_2)+2C\widetilde{\eta }_1^2}{(%
\widetilde{A}+(B_1+B_2)\widetilde{\eta }_1^2+C\widetilde{\eta }_1^4)^{1/2}}%
\right\} =0  \label{Min}
\end{equation}
\begin{eqnarray}
\widetilde{A}\widetilde{\eta }_1^2+\frac 12B_1\widetilde{\eta }_1^4+\frac
13C%
\widetilde{\eta }_1^6+2n\left\{ (\widetilde{A}+3B_1\widetilde{\eta }_1^2+5C%
\widetilde{\eta }_1^4)^{1/2}+\right.  \nonumber \\ \left.+(\widetilde{A}+(B_1+B_2)\widetilde{\eta }_1^2+C%
\widetilde{\eta }_1^4)^{1/2}\right\}=4n\widetilde{A}^{1/2} \label{Cont}
\end{eqnarray}
where as before $n=NH^2/8\pi D^{3/2}$, $\widetilde{A}$ and $\widetilde{\eta }%
_1^2$ are the values corresponding to the first-order phase transition
temperature.

The applicability of these equations at the first-order phase transition for
the non-symmetric phase in the limit $B_2\rightarrow 0$ (to try to describe
the N-IC phase transition), implies: 
\begin{equation}
n\frac{C^{3/2}}{B_1^{1/2}B_2^{3/2}}<1  \label{criterio}
\end{equation}
revealing only the dependence on thermodynamic potential coefficients. It is
obvious that this condition is violated if $B_2$ is sufficiently small, and
therefore we can approach but not reach the N-IC phase transition.

In the same way as before the changes produced by defects in $\widetilde{A}$
and $\widetilde{\eta }_1^2$ (Eqs. \ref{basicoa} and \ref{basicoeta}), now
for a two component order parameter potential, are given in the linear
approximation by: 
\begin{equation}
\delta \widetilde{A}=-4n\sqrt{\frac{B_2C}{3\mid B_1\mid }}
\end{equation}
\begin{equation}
\delta \widetilde{\eta }_1^2=-2n\sqrt{\frac C{3B_1^2}}\left\{ 9-2\sqrt{%
\frac{B_2}{\mid B_1\mid }}+\sqrt{\frac{\mid B_1\mid }{B_2}}\right\}
\end{equation}
The applicability criterion of this linear approximation to Eqs. \ref{Min}
and \ref{Cont} is found easily imposing the natural condition $\delta 
\widetilde{\eta }_1^2<\widetilde{\eta }_1^2$. In the limit $B_2\rightarrow 0$: 
\begin{equation}
\delta \widetilde{A}=-4n\sqrt{\frac{B_2C}{3\mid B_1\mid }}  \label{deltaA2}
\end{equation}
\begin{equation}
\delta \widetilde{\eta }_1^2=-2n\sqrt{\frac C{3\mid B_1\mid B_2}}
\label{deltaeta2}
\end{equation}
and without the numerical factor, his applicability criterion is: 
\begin{equation}
n\frac{C^{3/2}}{B_1^{3/2}B_2^{1/2}}<1
\end{equation}
In this case, this applicability criterion is less restrictive than Eq. \ref
{criterio}. This means that one can, in principle, go beyond the linear
approximation remaining within the region of applicability of Eqs. \ref{Min}
and \ref{Cont} but it is not necessary for our discussion.

As we can see from equations \ref{deltaA2} and \ref{deltaeta2}, for an approach a N-IC transition, the
correction to the transition temperature goes to zero while the correction
to the order parameter discontinuity is negative and diverges. This means,
of course, that the perturbation theory is no more applicable here. But
also indicate that while defects suppress very strongly the order
parameter discontinuity they do not influence (or weakly influence) the
phase transition temperature.

\section{Approaching a Normal-Incommensurate Transition: defects and
critical fluctuations}

In the preceding section we have considered phase transitions that are
first-order irrespectively to the order parameter fluctuations.
However there exist another type of first-order
phase transitions: those that are discontinuous due to critical
fluctuations. In Sec.II we shown that structural N-IC phase
transitions belongs to this class if it takes place in a broad vicinity of
the mean-field tricritical point. In this section we study influence of
defects on this type of transitions. 

We study a N-C
phase transition of the same symmetry class that in Sec.IV and
we follow the evolution of the results when a N-IC phase transition is
approached. For the thermodynamic potential one has now:  
\begin{eqnarray}
\Phi ({\boldsymbol \eta }_0,...,{\boldsymbol \eta }_{{\mathbf k}},...) &=&\frac 12A\eta
_1^2+\frac 14\widetilde{B}_1\eta _1^4+\frac 16C\eta _1^6-\nonumber \\  &&-m\left\{(A+3\beta
\eta _1^2+5C\eta _1^4)^{3/2}+ (A+(\widetilde{B}_1+B_2)\eta _1^2+C\eta
_1^4)^{3/2}\right\}+\nonumber \\&&+{\sum_{{\mathbf k}} }\frac 12\chi _1^{-1}(\eta
_1,k)\eta _{1{\mathbf k}}\eta _{1-{\mathbf k}}+{\sum_{{\mathbf k}} }\frac 12\chi
_2^{-1}(\eta _1, k)\eta _{2{\mathbf k}}\eta _{2-{\mathbf k}}
\end{eqnarray}
where as before $m= T/12\pi D^{3/2}$, but now making the definitions:\begin{equation}
\widetilde{B_1}\equiv B_1-\frac{2r^2}K
\end{equation}
\begin{equation}
\beta \equiv B_1-\frac{2r^2}{3K}-\frac{4r^2}{3\gamma }\equiv
\widetilde{B_1}+\Delta 
\end{equation} 
The coefficients
\begin{eqnarray}
\chi _1^{-1}(\eta _1,k) &\equiv &A+3\beta \eta _1^2+5C\eta _1^4+Dk^2 
-3m\left\{ (A+3\beta \eta _1^2+5C\eta _1^4)^{1/2}(3\beta
+10C\eta _1^2)+\right. \nonumber \\
&&\left. +(A+(\widetilde{B}_1+B_2)\eta _1^2+C\eta _1^4)^{1/2}(%
\widetilde{B}_1+B_2+2C\eta _1^2)\right\} 
\end{eqnarray}
\begin{eqnarray}
\chi _2^{-1}(\eta _1,k) &\equiv &A+(\widetilde{B}_1+B_2)\eta _1^2+C\eta
_1^4+Dk^2 -3m\left\{ (A+3\beta \eta _1^2+5C\eta _1^4)^{1/2}(3\beta
+10C\eta _1^2)+\right. \nonumber \\
&&\left. +(A+(\widetilde{B}_1+B_2)\eta _1^2+C\eta _1^4)^{1/2}(%
\widetilde{B}_1+B_2+2C\eta _1^2)\right\} 
\end{eqnarray}
are calculated as in Sec.II performing the integration in Eq. \ref{integration},
but now keeping out of this integration ${\boldsymbol \eta }_0$ and ${\boldsymbol \eta 
}_{{\mathbf k}^{\prime }}$ where ${\mathbf k}^{\prime }$ is a fixed wave vector.  

Defects are considered adding to this potential the term: 
\begin{equation}
-{\sum_{{\mathbf k}} }(h_{1{\mathbf k}}\eta _{1-{\mathbf k}}+h_{2{\mathbf k}}\eta _{2-{\mathbf k}})
\end{equation}

Proceeding as before with this thermodynamic potential we obtain in the limit $B_2\rightarrow 0$ :
\begin{equation}
\delta A^ \prime=-2n\left( \frac{2C}{3m}\right)^{1/3}
\end{equation}
\begin{equation}
\delta \eta _1^{\prime2}=-n\left(
\frac{2C}{3m}\right)^{1/3}\left(\frac
{1}{3\beta B_2}\right)^{1/2}
\end{equation}
close to the mean-field tricritical point. We see that for an approach a N-IC
transition, the correction to the order parameter discontinuity is negative
and diverges in the same way that it do for the Landau theory (Sec. IV).
Here the correction to the transition
temperature also remains finite but do not tends to zero. As well as in the preceding section this
indicates that while the defects suppress very strongly the order parameter
discontinuity even for small defect concentrations but they weakly influence
the phase transition temperature.

\section{Conclusions}

Hypothetical N-IC first-order phase transitions are strikingly different from
other ''ordinary'' first-order phase transitions where phase temperature
always diminishes because of influence of defects, see Fig.\ref{figure}. Lowering of
transition temperature in Fig.\ref{figure} (a), (b) is consistent with the expectation that
at some high enough defects concentration the phase transition temperature
becomes equal to zero and the phase transition no more exists. The practical
absence of this shift for a N-IC first-order phase transition (Fig.\ref{figure} (c))
seems to indicate that the crystal with defects "remembers" the first-order transition temperature in the pure material. This temperature might correspond to fairly rapid but still continuous changes in the satellite intensities with the temperature that might be interpreted as second-order transition. The results of this paper emphasize once more the strong influence of defects in the properties of incommensurate phases. 

We would like to thank S.A. Minyukov, J.J. S\'aenz and the members of his
group specially to A. Garc\'\i a-Martin and L.S. Froufe for useful
discussions. Special thanks to D. Sanchez-Soria for her help and support.

\begin{figure}
\caption{(a)Influence of "random temperature" defects on the phase diagram
of a first-order phase transition with elasticity taken
account. (b)Influence of "RLF" defects on the phase diagram of a first-order phase transition with
elasticity taken account. If the elasticity is neglected the same behaviour
that in (a) is obtained. (c)Expected influence of "RLF" defects on the
phase diagram of a N-IC first-order
phase transition}
\label{figure}
\end{figure}

\begin{thebibliography}{99}
\bibitem{nota} Numerical coefficients of this results are different respect to
their values in Ref.\onlinecite{Lev99}
\end{thebibliography}
\end{document}